\documentstyle[11pt,amssymb]{article}

\makeatletter
\@addtoreset{equation}{section}
\makeatother

\def\dalemb#1#2{{\vbox{\hrule height .#2pt
        \hbox{\vrule width.#2pt height#1pt \kern#1pt
                \vrule width.#2pt}
        \hrule height.#2pt}}}

\def\cao{{\cal O}}

\def\cn{{\cal N}}

\def\0{{\sst{(0)}}}
\def\1{{\sst{(1)}}}
\def\2{{\sst{(2)}}}
\def\3{{\sst{(3)}}}
\def\4{{\sst{(4)}}}
\def\5{{\sst{(5)}}}
\def\6{{\sst{(6)}}}
\def\7{{\sst{(7)}}}
\def\8{{\sst{(8)}}}
\def\n{{\sst{(n)}}}

\def\half{{\textstyle{1\over2}}}

\def\qu{{\textstyle{1\over 4}}}

\let\a=\alpha  \let\g=\gamma \let\d=\delta \let\e=\epsilon
  \let\q=\theta  
\let\l=\lambda \let\m=\mu \let\n=\nu  \let\r=\rho
 \let\t=\tau    
    \let\L=\Lambda
    
 \let\W=\Omega   
\let\la=\label  
  
\def\nn{\nonumber} \def\bd{\begin{document}} \def\ed{\end{document}}
\def\ds{\documentstyle} \let\fr=\frac \let\bl=\bigl \let\br=\bigr
\let\Br=\Bigr \let\Bl=\Bigl
\let\bm=\bibitem
\let\na=\nabla
\let\pa=\partial \let\ov=\overline
\newcommand{\be}{\begin{equation}}
\newcommand{\ee}{\end{equation}}
\def\ba{\begin{array}}
\def\ea{\end{array}}
\def\ft#1#2{{\textstyle{{\scriptstyle #1}\over {\scriptstyle #2}}}}
\def\fft#1#2{{#1 \over #2}}
\def\del{\partial}
\def\sst#1{{\scriptscriptstyle #1}}
 \def\oneone{\rlap 1\mkern4mu{\rm l}}
\def\ie{{\it i.e.\ }}
\def\via{{\it via}}
\def\semi{{\ltimes}}
\def\str{{\rm str}}
\def\Dm{{{D_{\sst{max}}}}}
\def\vac{ \left | 0 \right \rangle }
\def\kvac{ \left | k \right \rangle }

\def\sp{\; \; \;}

\def\bol{ \left | B (p^+) \right \rangle}
\def\bo1{ \left | B^0 (p^+) \right \rangle}

\def\bolt{ \left | B (p^+) \right \rangle_{\t}}

\def\boxl{ \left | B (x^-) \right \rangle}

\def\<{ \langle }
\def\>{ \rangle }

\newcommand{\hsp}{\hspace{0.5cm}}

\newcommand{\ho}[1]{$\, ^{#1}$}
\newcommand{\hoch}[1]{$\, ^{#1}$}
\newcommand{\bea}{\begin{eqnarray}}
\newcommand{\eea}{\end{eqnarray}}
\newcommand{\ra}{\rightarrow}
\newcommand{\lra}{\longrightarrow}
\newcommand{\Lra}{\Leftrightarrow}
\newcommand{\ap}{\alpha^\prime}
\newcommand{\bp}{\tilde \beta^\prime}
\newcommand{\tr}{{\rm tr} }
\newcommand{\Tr}{{\rm Tr} }
\newcommand{\NP}{Nucl. Phys. }

\newcommand{\ams}{{\it Institute for Theoretical Physics,
University of Amsterdam, \\
Valckenierstraat 65, 1018XE Amsterdam, The Netherlands} \\
{\tt skenderi,taylor@science.uva.nl}}
\newcommand{\auth}{Kostas Skenderis and Marika Taylor}

\begin{document}
\begin{flushright}
\hfill{\bf hep-th/0604169}\\
\hfill{ITFA-2006-18}
\end{flushright}

\vspace{15pt}

\begin{center}

{\Large \bf Holographic Coulomb branch vevs}

\vspace{20pt}

{\Large \auth}
\vspace{15pt}

\vspace{8pt}

{\ams}

\vspace{15pt}

\underline{ABSTRACT}
\end{center}

We compute holographically the vevs of {\it all} chiral 
primary operators for supergravity solutions corresponding
to the Coulomb branch of $\cn=4$ SYM and find {\it exact}
agreement with the corresponding field theory computation.
Using the dictionary between 10d geometries and field theory
developed to extract these vevs, we propose
a gravity dual of a half supersymmetric deformation
of $\cn=4$ SYM (on $Mink_4$) by certain irrelevant operators.

\noindent

\pagebreak
\setcounter{page}{1}

\section{Introduction}

Since its formulation \cite{Maldacena:1997re}-\cite{W}
there have been numerous tests of
the AdS/CFT duality, see \cite{Aharony:1999ti,D'Hoker:2002aw} 
for reviews. Soon afterward
the duality was extended to non-conformal field theories obtained
by deformations or vevs of the original CFT. In such cases
the duality involves an asymptotically AdS spacetime. However,
quantitative tests of the correspondence in this more general set up 
are rather scarce. 

Perhaps one of the simplest cases to explore is that 
of the Coulomb branch of the $\cn=4$ SYM theory. The theory still
possesses 16 supercharges and supersymmetry protects the vevs 
from quantum corrections. This is thus an ideal case 
for testing gravity/gauge theory duality away from the 
conformal point. Although it has long been 
recognized that there is a one to one correspondence between 
(the near-horizon limit of) multicenter D3 brane
solutions and the CB of $\cn=4$ SYM \cite{Kraus:1998hv}
a precise gravitational computation of the vevs was never 
done (apart from in the specific cases reviewed below). These solutions 
are determined by a harmonic function and Klebanov and Witten 
proposed in \cite{Klebanov:1999tb} that the vevs can be 
extracted from the harmonic function. The results of this
paper confirm that expectation. 

More quantitative progress has been achieved for the specific 
case of a distribution of D3 branes on a disc. In this 
case there is an associated solution of the five dimensional
gauged supergravity \cite{Freedman:1999gk} and using the 
technology of holographic renormalization \cite{dHSS}  the 
expectation values (and 2-point functions) of the stress energy 
tensor and of a dimension two gauge invariant operator were 
computed and shown to agree with field theory expectations in 
\cite{BFS1,BFS2}.

The restriction to this subsector of operators was due to starting from 
the five dimensional supergravity solution. More recently in
\cite{Skenderis:2006uy} we have 
developed a precise holographic map that allows one also to
treat operators dual to fields that are not part of the five
dimensional gauged
supergravity. Applying this map to the corresponding ten dimensional 
supergravity solution we additionally computed the vev of 
a dimension 4 operator and again found exact agreement with field
theory. This was the first quantitative computation of a vev 
of an operator dual to field that is not part of the $5d$ gauged
supergravity.

In this letter we essentially solve the problem in its
most general form. We consider a point in the 
CB of $\cn=4$ SYM characterized in the large $N$ limit 
by a general distribution of eigenvalues and we show that the 
gravitationally computed vevs of {\it all} gauge invariant 
operators are in {\it exact} agreement with the field
theory answer. We should note, however, there is 
still an open technical issue regarding the 
cancellation of certain terms that was checked only 
for a specific case (see the discussion in section \ref{vevs}).

Finally we point out that one can include
non-normalizable terms in the
harmonic function appearing in the supergravity solution for separated
D3-branes and these can also be interpreted in AdS/CFT. 
Such terms correspond to deformations of the SYM
theory by half supersymmetric irrelevant operators of the form 
$\Tr (F^4 X^k)$. 

The paper is organized as follows. In the next section we
discuss the field theory side of the story. In section 
3 we review the Coulomb branch solutions; in section 4
we summarize the results from \cite{Skenderis:2006uy} whilst
section 5 contains the holographic computation of the 
vevs for the general case. In section 6 we propose 
a gravity dual for certain deformations of $\cn=4$ SYM preserving 
16 supercharges. Appendix A contains a proof of the 
addition theorem that is used in the extraction
of vevs in section 5.
 
\section{$\cn=4$ SYM on the Coulomb branch}

${\cal N}=4$ 
SYM contains 6 scalar fields $X^{i_1}$ in the adjoint representation
of the gauge group that we take to be
$SU(N)$. The Coulomb branch (CB) of ${\cal N}=4$ SYM corresponds to giving 
a vacuum expectation value (vev) to the scalars subject to the condition 
$[X^{i_1}, X^{i_2}]=0.$ A useful parametrization of the CB 
branch is in terms of vevs of composite operators.
The relevant operators here are the chiral primaries (CPOs),
\be \label{CPO}
\cao^{I_1} = {\cal N}_{I_1} C^{I_1}_{i_1 \cdots i_k} {\rm{Tr}}(X^{i_1}
\cdots X^{i_k}),
\ee
where $C^{I_1}$ is
a totally symmetric traceless rank $k$ tensor of $SO(6)$ which is
normalized such that $\left < C^{I_1} C^{I_2} \right > = C^{I_1}_{i_1
\cdots i_k} C^{I_2}_{i_1 \cdots i_k} = \d^{I_1 I_2}$ and 
${\cal N}_{I_1}$ is a normalization factor. We choose this factor
such that the normalization of the 2-point functions computed in field theory 
and in supergravity is the same
\be \label{norm}
{\cal N}_{I_1} = \frac{N}{\pi^2} 2^{k/2} (k-2) \sqrt{\frac{k-1}{k}}
\left(\frac{2 \pi^2}{\l}\right)^{k/2}
; \qquad k \neq 2,
\ee
where $\l$ is the 't Hooft coupling. Note the field theory computation of the 
2-point functions is done in the large $N$ limit and with 
canonically normalized scalars. The $k=2$ case is obtained by replacing 
the factor of $(k-2)$ by 2.

We now consider an arbitrary smooth unit 
normalized distribution of eigenvalues, $\r(x)$, where $x$
parametrizes $R^6$. In the large $N$ limit the trace may be evaluated
via such a continuous eigenvalue distribution
\be
\left < C ^{I}_{i_1 \cdots i_k} \rm{Tr} (X^{i_1} \cdots X^{i_k})
\right > = N \int d^6x \rho(x) (C^I_{i_1 \cdots i_k}
x^{i_1} \cdots x^{i_k} )
\ee
giving the following formula for the vevs
\be \label{vev}
\<\cao^{I_1}\> = \frac{N^2}{\pi^2} 2^{k/2} (k-2) \sqrt{\frac{k-1}{k}}
\left(\frac{2 \pi^2}{\l}\right)^{k/2} 
\int d^6x \rho(x) (C^I_{i_1 \cdots i_k}
x^{i_1} \cdots x^{i_k} ), \qquad k \neq 2,
\ee
where again for $k=2$ one replaces the factor of $(k-2)$ by 2.
The aim of this paper is to reproduce this formula from supergravity
for all $k$ and general eigenvalue distributions. 

Before proceeding let us briefly review the properties
of a uniform distribution of eigenvalues of $X^1$ and $X^2$
on a disc of radius $a$ and vanishing vev for the 
remaining scalars, $\<X^3\> = \<X^4\>=\<X^5\>=\<X^6\>=0$.
In this case the vevs break the R symmetry from 
$SO(6)$ to $SO(2) \times SO(4)$.
This example was recently discussed in some detail in \cite{Skenderis:2006uy}
and the corresponding vevs were shown to be equal to
\be \la{vev1}
\left < \cao^{2n} \right > = 
\frac{ {\cal N}_{2n} a^{2n}}{ 2^n  \sqrt{2n+1}} N
\ee
where the operators here are the singlets under the decomposition
of $SO(6)$ into $SO(2) \times SO(4)$.

\section{Coulomb branch solutions} \label{CBsection}

A generic Coulomb branch solution describing a 
distribution of D3-branes is given by:
\bea \la{cbs}
ds^2 &=& H(x_\perp) ^{-1/2} d x_{||}^2
+ H(x_\perp)^{1/2} d x_\perp^2 \\
F&=&\frac{1}{4} (d H^{-1} \wedge \omega_{||}
- *_\perp d_\perp H) \nonumber 
\eea
where $\omega_{||}$ is the volume form in the (flat) worldvolume directions
and
$*_\perp$ and $d_\perp$ refer to the Hodge star and exterior derivative
in the flat transverse directions and $H$ is a harmonic
function,
\be \label{harm}
\Box H =0.
\ee
For the solution to be asymptotically AdS, $H$ should behave to 
leading order as $r^{-4}$ when $r \to \infty$,  where $r$ is the radial
coordinate of the flat overall transverse direction. The most 
general solution of (\ref{harm}) with these boundary conditions is
\be \label{H_harm}
H = \sum_{k, I} h_{kI} \frac{Y_{k}^{I} (\vec{\theta}) }{r^{k+4}},
\ee
where the coordinates on the flat $R^6$ are $(r,\vec{\theta})$ with
$\vec{\theta}$ labeling the coordinates on the $S^5$. 
$Y_k^I$ is a normalized spherical
harmonic of degree $k$ with $I$ labeling its remaining quantum
numbers; normalized as 
\be
\int_{S^5} Y^{I_1}_{k_1} Y^{I_2}_{k_2} = \pi^3 \frac{\d^{I_1 I_2} \d_{k_1
    k_2}}{2^{k_1-1} (k_1+1) (k_1+2)} \equiv \d^{I_1 I_2} z_{k_1}.
\ee
The leading order term is given by 
$h_{00} = L^4 = 4 \pi g_s N (\a')^2$ if the total number of 
D3-branes is to be $N$, whilst by
measuring distances from the centre of mass one can as usual choose the
$k=1$ terms to vanish. We will discuss more general boundary conditions
for $H$ in section \ref{def}.

The harmonic function corresponding to a source distribution 
of D3-branes $\rho(x)$ (normalized to one) is given by
\be
H = L^4 \int d^6y \frac{\rho(y)}{ \left | x - y \right |^4},
\ee
We show in appendix \ref{addition_theorem} that the asymptotics of
this harmonic function can be written as in (\ref{H_harm}) with coefficients
\be \la{exp}
h_{kI} = 2^{k} (k+1) L^4 \int d^6x \rho(x) \left (C^I_{i_1 \cdots i_k}
x^{i_1} \cdots x^{i_k} \right )
\ee
where $C^I_{i_1 \cdots i_k}$ is totally symmetric and traceless and the
basis of $C^I$ is orthonormal. Comparing (\ref{exp}) with (\ref{vev})
we see that these coefficients are proportional to the vevs
of the CPOs. 

In the case of a uniform distribution of D3 branes 
on a disc of radius $l$  one can straightforwardly do the integral in 
(\ref{exp}) \cite{Skenderis:2006uy} with the result being
\be \label{h_disc}
h_{2n0} = L^4 2^n \sqrt{2 n +1} l^{2 n}
\ee
where $I=0$ signifies that only harmonics that are singlets under
the decomposition  $SO(6) \to SO(2) \times SO(4)$ are involved.

Finally let us comment on the case of a spherical shell of D3 branes of radius
$R$, for which the harmonic function is \cite{Kraus:1998hv}
\be
H = \frac{L^{4}}{r^4}, \hsp r \ge R; \hsp
H = \frac{L^4}{R^4}, \hsp r < R,
\ee
so the geometry is flat within the shell. The asymptotics of this
harmonic function are trivially of the form (\ref{H_harm}), with the
absence of any perturbation relative to AdS corresponding to the fact
that there are no vevs for CPOs, since there are no $SO(6)$ singlet
CPOs. The interpretation of constant terms in the harmonic function
will be discussed in section 6.

\section{General method for extracting vevs}

In this section we will give a brief review of the methods developed
in \cite{Skenderis:2006uy} for extracting vevs from a given
asymptotically $AdS_5 \times S^5$ geometry. The first step is 
to write the solution as the $AdS_5 \times S^5$ solution plus
a deviation\footnote{We follow the conventions
of \cite{Skenderis:2006uy}. Our index conventions are:
$M,N,...$ are $10d$ indices, $\m,\n,...$ are $AdS_5$ indices,
$a,b,...$ are $S^5$ indices. $x$ denotes AdS coordinates and $y$ 
$S^5$ coordinates.}
\bea
g_{MN} &=& g^o_{MN} + h_{MN}, \\
F_{MNPQR} &=&  F^o_{MNPQR} + f_{MNPQR}.\nn
\eea
where 
$g^o_{MN}$ and $F^o_{MNPQR}$ are the metric and 5-form of 
the $AdS_5 \times S^5$ solution. (The solutions under consideration
here do not involve the other supergravity fields, so we restrict our
discussion to the metric and five form.) Next we expand the perturbations
$h_{MN}, f_{MNPQR}$ in $S^5$ harmonics. The general expansion 
is given in \cite{Skenderis:2006uy}; here we only quote the 
following two components since these will be useful later:
\bea \label{harm_exp}
h_{a}^a(x,y) &=& \sum \pi^{I_1}(x) Y^{I_1}(y) \\
f_{a b c d e}(x,y) &=& \sum b^{I_1}(x) \L^{I_1} \e_{abcde} Y^{I_1}(y) 
\eea
where $Y^{I_1}$ are scalar spherical harmonics and $\L^{I_1}$ 
is the eigenvalue of the scalar harmonic under (minus) the d'Alembertian. 
It is convenient to introduce the following linear combinations of  
$\pi^{I_1}$ and $b^{I_1}$
\be \la{l1}
s^{I_1} = \frac{1}{20 (k+2)} (\pi^{I_1} - 10 (k+4) b^{I_1}), \qquad
t^{I_1} = \frac{1}{20 (k+2)} (\pi^{I_1} + 10 k b^{I_1}),
\ee
since these combinations are mass eigenstates of the linearized field equations
around $AdS_5 \times S^5$ \cite{Kim:1985ez}. The $s^{I_1}$ fields
correspond to the chiral primary operators in (\ref{CPO}) and the 
$t^{I_1}$ fields to half supersymmetric operators of the 
schematic form $\Tr F^4 X^k$.

There are three ingredients that enter into the map from 
coefficients of the harmonic expansion in (\ref{harm_exp}) to
vevs in the dual QFT. The first is the construction
of gauge invariant variables, if the solution is not
in de Donder gauge (which indeed is often not a convenient gauge choice for
the asymptotic expansion).
De Donder gauge, $D^a h_{(ab)}=D^a h_{a\m}=0$, means that
the harmonic expansion of the metric deviations does not involve 
terms with derivatives of the harmonics. It is easy to see that 
the CB metrics (\ref{cbs}) in the coordinate system
where (\ref{H_harm}) holds satisfy this requirement
(but notice that the CB solutions expressed in the coordinates of
\cite{Freedman:1999gk} are not in de Donder gauge, 
as discussed in \cite{Skenderis:2006uy}).
So since we can conveniently work in de Donder gauge here,
we do not need to review the construction of gauge invariant
variables (and we simplify our notation relative to \cite{Skenderis:2006uy}
by dropping the tildes and hats from our notation).  

The second ingredient is the non-linear KK map from ten dimensional fields 
to five 
dimensional ones. For simplicity let us restrict to the subsector
involving $s^2$ and $s^4$ fields that are singlets
under $SO(6) \to SO(2) \times SO(4)$. The non-linear reduction map to 
second order in the fields reads \cite{LMRS} 
\bea \label{KKmap}
S^2 &=& \frac{\sqrt{8}}{3}\left(s^2 - \frac{2 \sqrt{3}}{15}  (s^2)^2 
-\frac{1}{12 \sqrt{12}} D_\m s^2 D^\mu s^2 \right) \\
S^4 &=& \frac{2 \sqrt{3}}{5}\left(s^4 - \frac{83}{18 \sqrt{5}} (s^2)^2 
-\frac{7}{18 \sqrt{5}} D_\m s^2 D^\mu s^2\right) \nonumber
\eea
where capital letters denote five dimensional fields
and small letters ten dimensional fields. The fields 
$S^2$ and $S^4$ defined by (\ref{KKmap}) solve the five dimensional
equations up to second order
\be \label{5deqn}
\Box S^2 = -\frac{4}{\sqrt{6}} (S^2)^2, \qquad \Box S^4 = 0.
\ee
In (\ref{KKmap})-(\ref{5deqn}) we have included only terms that can potentially
contribute to the asymptotic expansion of $S^2$ and $S^4$ up to 
the order required for extraction of vevs. More generally to extract
the vev of the operator dual to $S^{\Delta}$ one would need to retain
terms involving up to $\Delta/2$ fields. The overall field 
normalization is chosen such that in the 
$5d$ action, all fields are canonically normalized,
apart from an overall factor of $N^2/2 \pi^2$.

The final step is to use the method of holographic renormalization 
to extract the vevs from the asymptotics of the $5d$ fields. 
This is by now a standard procedure (see \cite{Skenderis:2002wp} for a 
review), except that here one needs to include additional terms to 
accommodate extremal couplings (see section 5.4 of \cite{Skenderis:2006uy}).
The relation between field asymptotics and vevs is most transparent 
in Hamiltonian variables where the radius plays the 
role of time. The 1-point functions are then related to
the radial canonical momenta of the bulk fields \cite{PS}.
For the operators $\cao^2$ and $\cao^4$ (which are 
singlets under $SO(6) \to SO(2) \times SO(4)$) the relations are 
\cite{Skenderis:2006uy}:
\bea
\< \cao^{4} \> &=& \pi_{(2)}^2 \nn \\
\< \cao^{4} \> &=& \pi_{(4)}^4 
+ \frac{3 {\cal N}_{4}}{\sqrt{5} {\cal N}_{2}^2 N} (\pi^2_{(2)})^2 
\label{o4_p}
\eea
where $\pi_{(k)}^{m}$ indicates the part of the canonical momentum
of the field $S^m$ that scales with weight $k$
and $\cn_{k}$ are the normalization factors given in (\ref{norm}). 
The relevant part of the canonical momenta can be expressed in terms 
of the asymptotic expansion of the $5d$ fields as follows
\be \label{mom}
\pi^k_{(2 k-4)} = \frac{N^2}{2 \pi^2} (2 k- 4) [S^k]_{k}
\ee
where the notation $[A]_k$ indicates the coefficient of the $z^k$
term in $A$ and $z$ is the Fefferman-Graham radial coordinate.
The relation (\ref{mom}) holds for $k \neq 2$;  
when $k=2$ one should replaces the factor $(2 k-4)$ by 2.

For reasons which will become clear later
it is useful to express the vevs directly in terms of the 
coefficients that appear in the $10d$ solution. Using the results 
reviewed above one obtains
\bea
\< \cao^2 \> &=& \frac{N^2}{2 \pi^2} \frac{2 \sqrt{8}}{3} 
[s^2]_2 \\
\< \cao^4 \> &=& \frac{N^2}{2 \pi^2} \frac{4 \sqrt{3}}{5}
[2 s^4 
+ \frac{37}{9 \sqrt{5}} (s^2)^2 -\frac{7}{9 \sqrt{5}} (D s^2)^2]_4
\eea
The expression for $\< \cao^4 \>$ can be further simplified 
for solutions with $s^2$ only depending on the radial coordinate,
such as those under consideration in this paper. In such cases, 
$[(D s^2)^2]_4 = 4 [(s^2)^2]_4$ and we obtain
\be \label{o4}
\< \cao^4 \> = \frac{N^2}{2 \pi^2} \frac{4 \sqrt{3}}{5}
[2 s^4 + \frac{1}{\sqrt{5}} (s^2)^2]_4.
\ee
The general method outlined here 
can be applied to extract the vevs of all other 
higher dimension operators. However, the procedure becomes
complex as the operator dimension increases, since one has to retain
terms to higher and higher order in the reduction map, the field
equations and the relations which give the vevs. The purpose
of the current paper is to point out that simplifications occur for
the Coulomb branch solutions: for these
we will be able to carry out the vev computation for arbitrarily
high dimension operators.

\section{Obtaining the vevs} \label{vevs}

We now return to the Coulomb branch solutions and
express the harmonic function as 
\be
H = \frac{L^4}{r^4} + \delta H,
\ee
i.e. we separate the $k=0,I=0$ term in (\ref{H_harm}) that 
yields the $AdS_5 \times S^5$ part of the solution 
from the remaining terms. The metric and five form field become
\bea
ds^2 &=& L^2 r^2 (1 + \frac{r^4 \d H}{L^4})^{-\half}
dx_{||}^2 + L^2 (1 + \frac{r^4 \delta H}{L^4})^{\half}
(\frac{dr^2}{r^2} + d \Omega_{5}^2); \\
F_{5} &=& (L^4 - \frac{1}{4} r^5 \pa_{r} \delta H) d\W_5 + \qu r^3 dr
\wedge \ast_{S^5} D_{a} \Delta H   
+ \qu d ( L^4 r^4 (1 + \frac{\delta H r^4}{L^4})^{-1})
\wedge \omega_{||}. \nn
\eea
Note that the coordinates on $R^{3,1}$ have been rescaled as
\be \label{resc}
x_{||} \to L^2 x_{||}
\ee
so that the metric scales as $L^2$.

This coordinate system is manifestly compatible with the de Donder
gauge for fluctuations; we can immediately read off the following
expression for the field $b^I_k$ in (\ref{harm_exp}) as 
\be
b^{I}_k =  - \frac{h_{kI}}{4 L^4 k r^k}
\ee
whilst the trace of the fluctuation on the sphere is
\be
\pi = 5 \left ( (1 + \frac{r^4 \delta H}{L^4})^{1/2} - 1 \right).
\ee
Retaining only linear terms in the expansion of the square root gives 
\be \label{pi_lin}
\pi^{I}_k = \frac{5 h_{kI}}{2 L^4 r^k}.
\ee
Note however that there is no justification generically for retaining
only the linear term. Indeed the quadratic correction gives
\be \label{corr}
- \frac{5}{8 z(k)} \sum_{k_1, I_1, I_2} \frac{1}{L^8 r^k} h_{k_1 I_1}
h_{(k-k_1) I_2} \left < Y^{I_1}_{k_1} Y^{I_2}_{(k-k_1)} Y^{I}_{k}
\right >,
\ee
where the term in parenthesis indicates the triple overlap of the
scalar harmonics. This term is not suppressed compared to the linear term 
(recall the $h_{kI}$ contain a factor of $L^4$ so the factors of
$L$ cancel in (\ref{corr})); there is no small parameter in general.
The issue of non-liner terms first appears for dimension 4 operators. In this case
we constructed the holographic map from first principles
in \cite{Skenderis:2006uy}, as reviewed in the previous section,
so we should be able to understand whether the quadratic
corrections contribute. Notice that since $h_{kI}$ is proportional 
to the vev of a chiral primary operator, the quadratic correction
looks like a double trace contribution.

To understand this issue let us restrict to the special case of a disc
distribution which was the case understood in detail in 
\cite{Skenderis:2006uy}. In this case $s^2$ and $s^4$ are given by
\be
s^2 = \frac{1}{8} h_{20}, \qquad
s^4 = \frac{1}{16} h_{40} -\frac{1}{128 \sqrt{5}} (h_{20})^2,
\ee
where $h_{20}$ and $h_{40}$ are given in (\ref{h_disc}). The term 
in $s^4$ quadratic in  $h_{20}$ comes from (\ref{corr}). Inserting these
values in the expression for the vevs (\ref{o4}) we find that 
the  $(h_{20})^2$ part of $s^4$ is
precisely cancelled by the $(s_2)^2$ term! The remaining terms
are exactly right to reproduce (\ref{vev1}). Notice that the 
$(s_2)^2$ terms in (\ref{o4}) originate from two sources.
One is the non-linear terms in the KK map and the second are the 
non-linear terms in the 1-point function (\ref{o4_p}). Based on this result
we conjecture that the same type of cancellation will occur 
for all operators in the general case.

So now let us return to the general case and assume that such a cancellation takes 
place. We thus retain only the linear term (\ref{pi_lin})
and ignore non-linear terms in the KK map and non-linear 
terms in the 1-point function. The 10d fluctuations under these
assumptions are
\be
s^{I}_{k} = \frac{h_{kI}}{4 k L^4 r^k}, \qquad t^I_k=0,
\ee
so the canonically normalized five dimensional fields are equal to
\be
S^{I}_{k} = \frac{\sqrt{(k-1)}}{2^{k/2} \sqrt{k} (k+1)} \frac{h_{kI}}{L^4 r^k},
\qquad T^I_k=0,
\ee
where for completeness we also quote the values of the $t$ field.
This implies the following expression for the vev
\be
\left < {\cao}_{k}^I \right > = 
\frac{N^2}{2 \pi^2} (2k - 4) [S^{I}]_{k} 
= \frac{N^2}{\pi^2} 
\frac{(k-2) \sqrt{(k-1)}}{2^{k/2} \sqrt{k} (k+1)} \frac{h_{kI}}{L^4}, 
\ee
for $k \neq 2 $ whilst the formula for $k=2$ follows using the
replacement $(k-2) \rightarrow 2$. Using the identity (\ref{exp}) 
then gives a final expression for the vev as 
\be \label{sugra}
\left < {\cao}_{k}^I \right > = \frac{N^2}{\pi^2} 2^{k/2} (k-2)
  \sqrt{\frac{k-1}{k}} \int d^6x' \rho(x') (C^I_{i_1 \cdots i_k}
(x')^{i_1} \cdots (x')^{i_k} ),
\ee
for $k \neq 2$ with the analogous expression for $k=2$ being obtained
by the replacement $(k-2) \rightarrow 2$. To match supergravity 
and field theory normalizations, field theory lengths
must be scaled by a factor of $\sqrt{\l/2 \pi^2}$.
This is due to the rescaling in (\ref{resc}) and is explained in 
section 6.3 of \cite{Skenderis:2006uy}. Taking this factor into
account we find exact agreement between the supergravity
and field theory computations in  
(\ref{sugra}) and (\ref{vev})!

This result is highly non-trivial: whilst all the vevs must necessarily 
be encoded in the harmonic function, it is surprising that one can
extract them so simply by the above linearization of the ten-dimensional 
fields. Note that this procedure for extracting the Coulomb branch
vevs was suggested in \cite{Klebanov:1999tb}, although the
interpretation of the neglected terms was and remains unclear.
Presumably the linearization can be rigorously justified in this case by
proving that the non-linear terms cancel when one carries out the
holographic renormalization procedure directly in ten dimensions. This
issue is currently under investigation. 

One would also not anticipate that there is an analogous route for
extracting vevs from general asymptotically $AdS_5 \times S^5$
solutions. In the CB example, the simplifications arose when we
expanded the solution in a particular radial coordinate, which turned
out to be exactly the Fefferman-Graham coordinate, and then
linearized in a natural way. Had we expanded the same solutions in the coordinate
systems of \cite{Freedman:1999gk}, we would have had no motivation for
retaining the required ``linearized'' subset of terms in the
perturbations. Indeed, in the discussion of the disk distribution in
\cite{Skenderis:2006uy}, only the complete gauge invariant fields $\pi^{I}_k$
played a role. For more general supergravity solutions, even those such as the
bubbling geometries of \cite{Lin:2004nb} which are built from harmonic
functions, it is not obvious how one would even define the above procedure
or why it should be justified. By contrast, the method for extracting
vevs presented in
\cite{Skenderis:2006uy} applies to {\it all} asymptotically $AdS_5 \times S^5$ solutions.

\section{Duals of 1/2-susy deformations of $\cn=4$ SYM} \label{def}

In section \ref{CBsection} we discussed the supergravity 
solution (\ref{cbs}) describing a distribution of D3-branes. The  
harmonic function $H$ entering the solution was
constrained to vanish as $r^{-4}$ in order for the 
solution to be asymptotically AdS. We discuss here 
solutions obtained by relaxing this condition, and their
interpretation using the AdS/CFT dictionary, in particular using the map
between ten dimensional fields and the gauge theory discussed in the
previous section. 

The most general solution of (\ref{harm}) is
\be \label{genh}
H = \sum_{k, I} \left(l_{k I} r^{k} + \frac{h_{k I}}{r^{k+4}}\right) Y^I_k 
\ee
The case with only the $k=I=0$ terms is D3 brane solution with the asymptotically
flat region included. The more general solution (\ref{genh})
still preserves 16 supercharges, and as we will shortly see, surprisingly
admits an AdS/CFT interpretation despite the fact that the solution
is not asymptotically AdS (in any usual usage of the word).

We proceed as in the previous section by writing
\be
H = \frac{L^4}{r^4} + \delta H
\ee
and keeping only the linear term in $\delta H$. The terms proportional 
to $h_{k I}$ have been discussed already, so we only 
consider here the new terms proportional to $l_{k I}$. From the
metric and five-form we read off
\be
\pi_k^I = \frac{5l_{kI}}{2 L^4} r^{k+4}, \qquad  
b_k^I= \frac{l_{kI}}{4 (k+4) L^4} r^{k+4}. 
\ee
Forming the $s$ and $t$ combinations we obtain
\be
s^I_k=0, \qquad t_k^I = \frac{l_{kI}}{4 (k+4)} r^{k+4}.
\ee
The radial behavior of the $t$ field is exactly right for the 
solution to correspond to a deformation of the $N=4$ SYM theory by 
the operators dual to the $t^k$ fields, with the deformation
parameter being (proportional to) $l_{kI}$. The restriction of 
keeping only linear terms in $\d H$ may be justified by 
considering the deformation parameters $l_{kI}$ to be 
small (although the results may hold more generally
due to cancellations as in the case of $h_{kI}$).

Such a deformation was discussed in \cite{Intriligator:1999ai} 
where it was argued that the gauge beta function continues to vanish. 
This corresponds to the fact that the dilaton and axion are constant in 
the solutions we discuss. In the same paper, it was argued 
(extending earlier work \cite{Gubser:1998kv,Gubser:1998iu}) that the 
full D3-brane solution (with the asymptotic flat region included)
is dual to $N=4$ SYM deformed by a dimension 8 operator. This 
is precisely the operator dual to $t^0$. In this paper we 
generalize this proposal to the general case of a deformation
by all operators dual to $t^k$. It is easy to check that all symmetries 
(susy, R-symmetry) match between the supergravity and 
field theory descriptions. Notice that the operators $t^k$ 
are irrelevant so one expects a strong backreaction on the 
asymptotics of the dual geometry. It would be interesting 
to explore this case further, especially since the geometry
is far from being asymptotically AdS so one may learn
about how holography works in more general contexts.

\section*{Acknowledgments}

The authors are supported by NWO, KS via the Vernieuwingsimplus grant 
``Quantum gravity and particle physics'' and 
MMT via the Vidi grant ``Holography,
duality and time dependence in string theory''.

\appendix

\section{The addition theorem for $SO(6)$ spherical harmonics} \label{addition_theorem}

In this appendix we prove (\ref{exp}). The harmonic
function 
\be
H = L^4 \int d^6y \frac{\rho(y)}{ \left | x - y \right |^4},
\ee
can be expanded in powers of $r=\sqrt{x^2}$ as 
\bea
H &=& \frac{L^4}{r^4} 
\int d^6 y \rho(y) \sum_{n \geq 0} (-1)^n (n+1) \left ( \frac{
  y^2}{r^2} - 2 y \frac{ {\bf{\hat{x}} \cdot \bf{\hat{y}}}} {r} \right
)^n; \la{ing1} \\
&=& \frac{L^4}{r^4} \int d^6y \rho(y) 
\sum_{n \geq 0} \sum_{m=0}^{n} (-1)^{n+m} (n+1)
\frac{2^m n!}{m! (n-m)!} \frac{ y^{2n-m} \cos^m(\g)}{r^{2n-m}}, \nn
\eea
where $\bf{\hat{x}}$ and $\bf{\hat{y}}$ are unit vectors
and $\cos(\g) =  \bf{\hat{x}} \cdot  \bf{\hat{y}}$. 
Collecting together terms of the same radial power we obtain
\be \la{two}
\sum_{n \geq 0} \sum_{m=0}^{n} (-1)^{n+m} (n+1)
\frac{2^m n!}{m! (n-m)!} \frac{ y^{2n-m} \cos^m(\g)}{r^{2n-m}}
= \sum_{k  \geq 0} \frac{y^k}{r^k} C^{(2)}_k (\cos \g),
\ee
where $C^{(2)}_k (\cos \g)$ is a Gegenbauer polynomial 
satisfying the following differential equation
\be \la{difeq}
\left ( (1 - z^2) \pa_z^2 - 5z \pa_z + k(k+4) \right ) 
C^{(2)}_k (z) = 0.
\ee
Properties of the Gegenbauer polynomials are given in \cite{AS};
in particular they satisfy an orthogonality relation
\be
\int_{-1}^{1} dz (1 -z^2)^{3/2} C^{(2)}_k(z) C^{(2)}_l(z) = \frac{1}{8} \pi
(k+1) (k+3) \d_{kl}
\ee
and are such that
\be
C_k^{(2)}(1) = \frac{1}{6} (k+1)(k+2)(k+3).
\ee
Now introducing the following coordinates on the $S^5$ 
\be \la{coor5}
ds^2 = d\q^2 + \sin^2 \q d\Omega_4^2,
\ee
$SO(5)$ singlet spherical harmonics satisfy the differential equation
\be
\frac{1}{\sin^4 \q} \pa_{\q} (\sin^4 \q \pa_{\q}) Y_k = - k(k+4) Y_k,
\ee
which is the same equation as (\ref{difeq}) with $z = \cos
\q$. The canonically normalized spherical harmonics are therefore given by
\be
Y_{k} (\q) =  \frac{\sqrt{3}}{2^{k/2} (k+1) \sqrt{(k/2 +1)(k+3)}} 
  C^{(2)}_k (\cos \q) \equiv \l_k  C^{(2)}_k (\cos
  \q) ;  \la{p1} 
\ee
with the values on the axis $\q = 0$ being 
\be
Y_{k} (0) =  \frac{ \sqrt{(k/2+1) (k+3)}} {\sqrt{3} 2^{k/2}} \equiv
y_k . \la{p2} 
\ee
Note that spherical harmonics which
are not $SO(5)$ singlets take the form $Y^{p}_{k}(\q) Y_{pk}(\q_4)$
with  $\q_4$ coordinates on $S^4$ and
\be
\frac{1}{\sin^4 \q} \pa_{\q} (\sin^4 \q \pa_{\q}) Y^p_k -
\frac{p(p+3)}{\sin^2 \q}  Y^p_k= - k(k+4) Y^p_k,
\ee
where $p$ is the $SO(5)$ eigenvalue (on the $S^4$) with $p \le k$. The relevant
solutions are again expressed in terms of Gegenbauer polynomials as
\be
Y^p_k \propto \sin^p \q C^{(p+2)}_{k-p} (\cos \q),
\ee
Since $C^{(p+2)}_{k-p}(1)$ is non-vanishing and finite, the total harmonic is zero
when $\q=0$. Thus only the $SO(5)$ singlet spherical harmonics are
non-vanishing at $\q = 0$. 

Comparing (\ref{ing1}) and (\ref{two}) with (\ref{H_harm}) and
  (\ref{exp}) implies that the following
identity needs to be proved
\be \la{add}
C^{(2)}_k (\cos \g) =  2^{k} (k+1) \sum_{I} Y^{I}_{k} (\q^y_5)
Y^{I}_{k} (\q_5),
\ee
where we use the identity
\be
C^{I}_{i_1 \cdots i_k} y^{i_1} \cdots y^{i_k} = y^k Y^{I}_{k}
(\q^y_5),
\ee
with $(\q^y_5)$ the coordinates on the $y$ 5-sphere. Now (\ref{add}) is
the exact analogue of the addition theorem for spherical harmonics
of $SO(3)$ used in electromagnetism and can be proved in exactly the
same way \cite{Jackson}. First note that in the coordinate system (\ref{coor5}) on
the sphere
\be
\cos \g = \cos \q \cos \q_y + \sin \q \sin \q_y (\cos \g_{4}),
\ee
where $\g_{4}$ is the angle separating the vectors on the $S^4$. Thus
when $\q_y = 0$ (it lies on the ``axis'') $\cos \g = \cos \q$.
Since the $SO(5)$ singlet harmonic is the only harmonic at level $k$
which is non-vanishing on the axis (\ref{add}) collapses to 
\be
C^{(2)}_k (\cos \q) =  2^{k} (k+1) Y_{k} (0)
Y_{k} (\q),
\ee
an identity which is manifestly true because of (\ref{p1}) and
(\ref{p2}). 

Now consider rotating the axes so that $\q_y$ is no longer zero. Then
the function $C^{(2)}_k (\cos \g)$ still satisfies the
covariant version of (\ref{difeq}), namely 
\be
( \Box + k(k+4)) C^{(2)}_{k} (\cos \g) = 0,
\ee
where $\Box$ is the Laplacian on the $S^5$ with coordinates $\q_5$. In
other words, the function can always be expanded in spherical harmonics of
rank $k$ as
\be
 C^{(2)}_k (\cos \g) = \sum_{I} \a^{I}_{k}( \q^y_5) Y^{I}_{k}
 (\q_5),
\ee
where the coefficients are given by 
\be \la{r1}
\a^{I}_{k}( \q_5^y) = z_{k}^{-1} \int_{S^5} d\Omega_5
Y^{I}_{k} (\q_5) C^{(2)}_k (\cos \g).
\ee
However, a generic function can be expanded in terms of spherical
harmonics as 
\be
f(\q_5) = \sum_{k,I} \beta_{kI} Y^{I}_{k} (\q_5),
\ee
where
\be
\beta_{kI} = \frac{1}{z_k} \int_{S^5} d\Omega_5 f(\q_5) Y^I_k(\q_5),
\ee
and in particular for the $SO(5)$ singlet coefficients
\be
\beta_{k} = \frac{\l_k}{z_k} \int_{S^5} d\Omega_5 f(\q_5)
C^{(2)}_{k} (\cos \q),
\ee
so that $f(\q = 0) = \sum_{k} \beta_k y_k$. Then (\ref{r1}) is the
$SO(5)$ singlet coefficient in an expansion of the function $Y^{I}_{k}
(\q_5)/\l_k$ in a series of $Y^{I}_{k}(\g, \cdots)$ (i.e. with respect
to the rotated axis discussed earlier). One can thus read off the
coefficient (\ref{r1}) as 
\be
\a^{I}_{k}( \q_5^y) = \frac{1}{y_k \l_k} Y^{I}_{k} (\q_5(\g,
\cdots))_{\g = 0} = 2^{k} (k+1)  Y^{I}_{k} (\q_5^y),
\ee
since in the limit $\g \rightarrow 0$ the angles $(\q,\cdots)$ go over
into $(\q_y,\cdots)$. This completes the proof of (\ref{add}) and
hence of (\ref{exp}). Note that (\ref{add}) also implies a sum rule
for harmonics of the same degree:
\be
\sum_{I} (Y^{I}_{k}(\q_5))^2 = 2^{-k} (1 + k/2) (1+ k/3),
\ee
analogous to that sometimes used in electromagnetism.

\end{document}